\documentclass[%
 aip,
 amsmath,amssymb,
 reprint,%
]{revtex4-1}

\usepackage{graphicx}
\usepackage{dcolumn}
\usepackage{bm}
\usepackage{gensymb}
\usepackage{array}
\usepackage{tabularx}
\usepackage{booktabs}
\usepackage{multirow}
\usepackage{siunitx} 
\usepackage[utf8]{inputenc}
\usepackage[T1]{fontenc}
\usepackage{mathptmx}
\usepackage{etoolbox}
\usepackage{float}
\makeatletter
\def\@email#1#2{%
 \endgroup
 \patchcmd{\titleblock@produce}
  {\frontmatter@RRAPformat}
  {\frontmatter@RRAPformat{\produce@RRAP{*#1\href{mailto:#2}{#2}}}\frontmatter@RRAPformat}
  {}{}
}%
\makeatother
\begin{document}

\preprint{AIP/123-QED}

\title[Construction and Commissioning of a Compact Xenon Gas Recycling System]{Construction and Commissioning of a Compact Xenon Gas Recycling System for a High-Harmonic Generation VUV Frequency Comb}
\author{G. Holthoff$^{*}$}
\email{georg.holthoff@physik.uni-muenchen.de}
\affiliation{ 
These authors contributed equally}
\affiliation{ 
Faculty of Physics, Ludwig-Maximilians-Universität München,\\ Am Coulombwall 1, 85748 Garching bei München, Germany}

\author{J. Rebhan}
\affiliation{ 
These authors contributed equally}
\affiliation{ 
Faculty of Physics, Ludwig-Maximilians-Universität München,\\ Am Coulombwall 1, 85748 Garching bei München, Germany}
\author{T. Teuner}
\affiliation{ 
Faculty of Physics, Ludwig-Maximilians-Universität München,\\ Am Coulombwall 1, 85748 Garching bei München, Germany}
\author{L. Guth}
\affiliation{Max-Planck-Institut für Kernphysik, Saupfercheckweg 1, 69117 Heidelberg, Germany,\\ Heidelberg Graduate School for Physics}
\author{D. Moritz}
\affiliation{ 
Faculty of Physics, Ludwig-Maximilians-Universität München,\\ Am Coulombwall 1, 85748 Garching bei München, Germany}%
\author{T. Teschler}
\affiliation{ 
Faculty of Physics, Ludwig-Maximilians-Universität München,\\ Am Coulombwall 1, 85748 Garching bei München, Germany}%
\author{M. Wiesinger}
\affiliation{ 
Faculty of Physics, Ludwig-Maximilians-Universität München,\\ Am Coulombwall 1, 85748 Garching bei München, Germany}%
\author{J. R. Crespo López-Urrutia}
\affiliation{Max-Planck-Institut für Kernphysik, Saupfercheckweg 1, 69117 Heidelberg, Germany,\\ Heidelberg Graduate School for Physics}
\author{P. G. Thirolf}
\affiliation{ 
Faculty of Physics, Ludwig-Maximilians-Universität München,\\ Am Coulombwall 1, 85748 Garching bei München, Germany}

\date{\today}

\begin{abstract}
A compact, closed-loop xenon recycling system for application in a High-Harmonic Generation (HHG) Vacuum-UV (VUV) frequency comb laser systems is designed, constructed and commissioned. It consists of three main sections, a low-pressure section (LPS) pre-filtering the gas of particulates, water and hydrocarbons, then a compression station (CS), allowing for a variable pressure of up to 60 bar, and a high-pressure section (HPS) with more filtering, pressure control and monitoring.
The system proves to meet all design criteria in gas quality (the sum off all hydrocarbon contaminations up to 120\,amu and CO$_{2}$ being below 10$^{-8}$\,mbar) and recovery efficiency (>99.8$\%$ for all expected operational cases, neglecting possible geometric gas catcher losses in the HHG laser itself). The operation and results are presented and discussed. Furthermore, the system is compared to the performance of a similar system from which it is derived.
\end{abstract}

\maketitle

\section{Introduction}
High-harmonic generation (HHG) is a non-linear optical process that generates odd-order harmonic frequencies of a high-intensity infrared laser that impinges on a centrosymmetric jet of gas or non-linear solid-state media. 
The laser light generated by this process ranges from the Deep-Ultraviolet (DUV) and Vacuum-UV (VUV) into the soft-X-ray section of the spectrum \cite{2_4_0_Corkum_verlorerner_Nobel_prize_HHG}. Molecular or noble gases may be used, but for most applications, noble gases are more suitable as they are chemically inert and pose no direct threat to mirrors or other optical components. Furthermore, krypton and xenon offer the highest harmonic yield in the longer-wavelength UV range due to their low ionization potential \cite{2_6_4_HHG_Gases}. 

The xenon-gas recycling system presented in this work was constructed for a VUV-frequency comb to be used for the direct laser excitation of Thorium-229 nuclei to their isomeric state, Thorium-229m. The project aims to excite the transition in a single $^{229}$Th$^{3+}$ ion trapped in a Paul trap and sympathetically cooled with $^{88}$Sr$^{+}$.  When successfully excited, this transition provides a possible frequency standard for a so-called "nuclear-clock" with potential relative uncertainties of $10^{-19}$ or below~\cite{1_Campbell_10^-19}. An in depth overview over the experiment is given in Ref. \citenum{Moritz2025}.
The VUV light starts out as an ultra stable IR-frequency comb centered at $\approx$1040\,nm. This comb light is then amplified and injected into an enhancement resonator, where the 7$^{th}$ harmonic output light at $\approx$148\,nm is generated through HHG \cite{VUVTh}. The resonator is treated in Ref. \citenum{Wissenberg2025}.
Operating HHG-laser systems with krypton or especially xenon is very costly: These gases are scarce on earth and only few companies worldwide extract them; prices are high and volatile. Under typical operational conditions for the laser used in the group, six-digit yearly operating costs just for gas consumption can therefore be expected without gas recovery.
A more economical solution is recycling, which must fulfill two essential criteria: Firstly, the enclosed noble gas must remain highly purified, as contaminants like hydrocarbons can lead to mirror degradation and thus a decrease in laser power ~\cite{2_6_3_Hydrocarbons_soleil,2_6_3_Hydrocarbons_VUV_Monolayer}. Secondly, the system must be as leak-tight as possible to avoid any loss of the valuable xenon gas.
Since the freezing point of xenon lies at $-$118~$^{\circ}$C, which is lower than most hydrocarbons, cryoextraction, i.e. freezing-out of other components or contaminants, may be an accessible solution \cite{2_6_3_Melting_Points}. To return to operating pressure this then either necessitates a compression stage, or liquefaction of the xenon at lower temperatures to generate bottle pressure after the gas evaporates again. 
Furthermore, HHG is often performed using a helium/xenon mixture to enhance gas flow speed and prevent steady state plasma formation \cite{Porat2018_PhaseMatched_XUV_Comb}. This is the case for the laser the system is designed for.
Therefore, this approach is again impractical as the helium does not freeze, and at the measured gas flows (6l of Xenon per hour, Helium added N times of this value, depending on the mixture, 1:4 or 1:9 are likely), even the cost incurred from the lost high-purity helium will be extensive over time, not to speak of the complexity and cost of a high-accuracy gas mixing system to return to the correct mixture.
Thus recompression and filtering at room temperature is the option deemed most stable and reasonable.
A recycling setup based on this closed-loop concept was developed by the group of J. Crespo at the MPIK in Heidelberg as described in Ref. \citenum{1_Griessbach_Paper}. Building upon this precursor design an adapted and modified version was designed, constructed and commissioned. The changes made to the design relative to the MPIK reference were chosen for the most efficient gas recovery and improved system compactness without compromising HHG output power or inflating component costs.

In the following, a brief setup overview is given, followed by a discussion of the key parameters relevant for characterizing the recycling system motivating the choice of conducted experiments and measurements. At the end the corresponding results are presented and discussed.

\section{Setup Overview}

\begin{figure*}[htbp]
    \centering
    \includegraphics[width=0.85\linewidth]{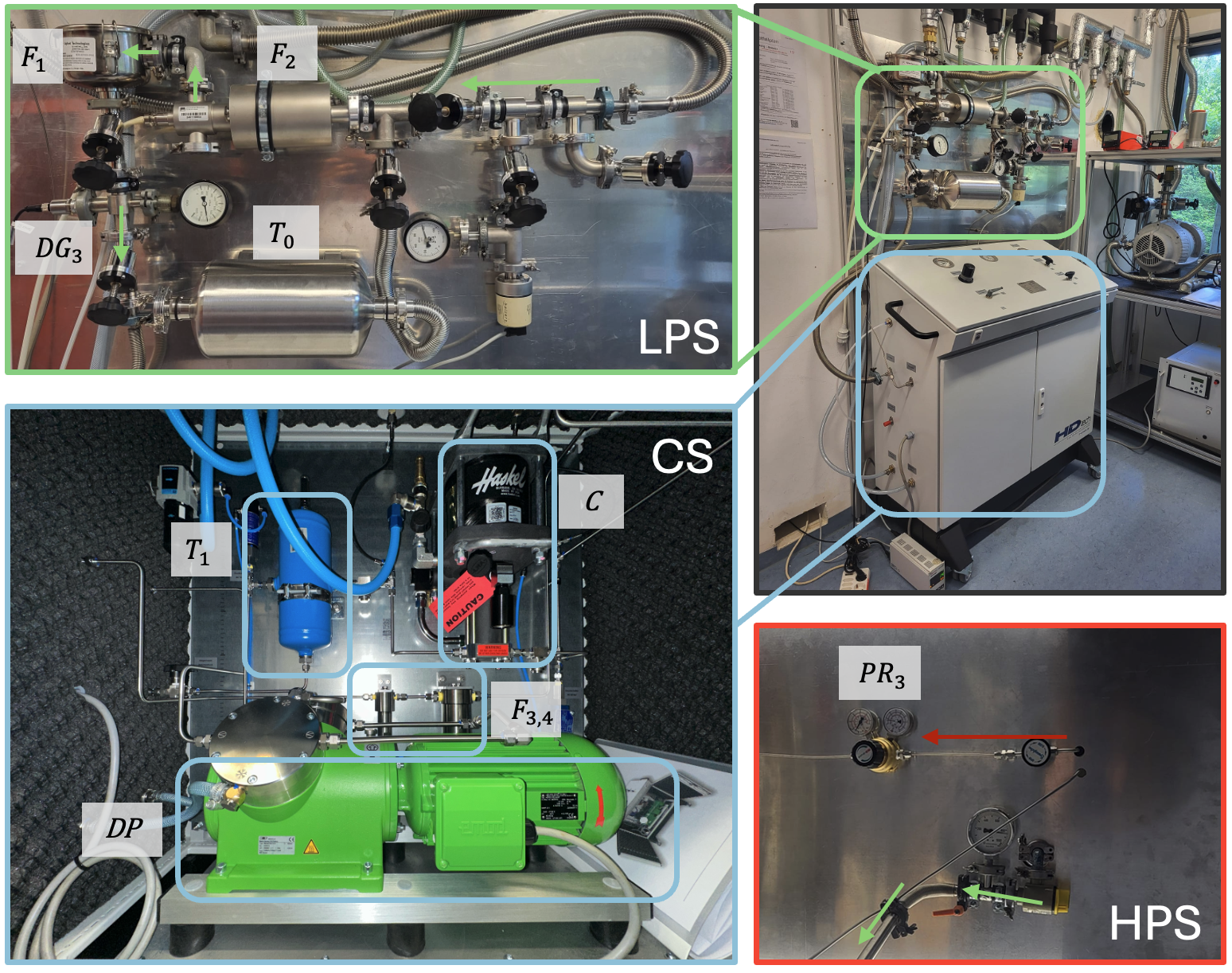}
    
    \vspace{0.2cm}
    
    \includegraphics[width=0.7\linewidth]{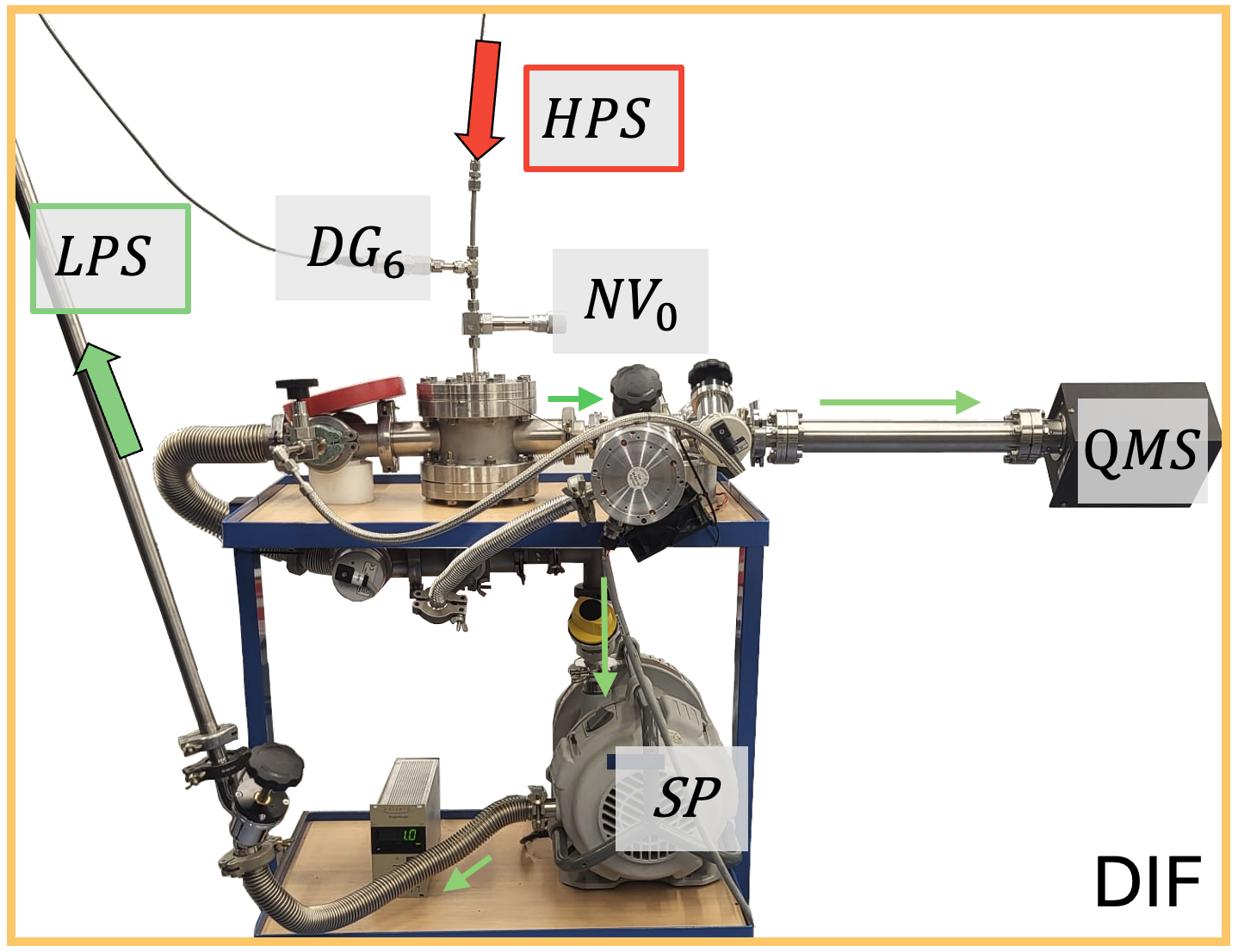}
    
    \caption{Photographs of the xenon recycling setup. Each section is outlined in a different colour: High pressure system (HPS)~-~red, low pressure system (LPS)~-~green, compressor station (CS)~-~blue and the ancillary differential pumping section for mass spectroscopy (DIF) in orange. DG$_{x}$ are digital pressure gauges, F$_{x}$ are the gas filters, T$_{x}$ are expansion tanks, PR$_{x}$ are pressure regulators and DP is the first compression stage diaphragm pump, concluding the main system. In the differential pumping section DIF, NV$_{0}$ is the main needle valve controlling the gas flow to the nozzle, QMS is a quadrupole mass spectrometer and SP the scroll pump evacuating the chamber and feeding the gas back toward the diaphragm pump at the input of the compression section. For a detailed listing of components see Table~\ref{tab:3.1.all_components}. The red arrow indicates gas flow at high pressure from the HPS, green arrows low pressure gas flow toward the LPS.}
    \label{fig:Setup_getrennt}
\end{figure*}

\begin{figure*}
    \centering
    \includegraphics[width=1\linewidth]{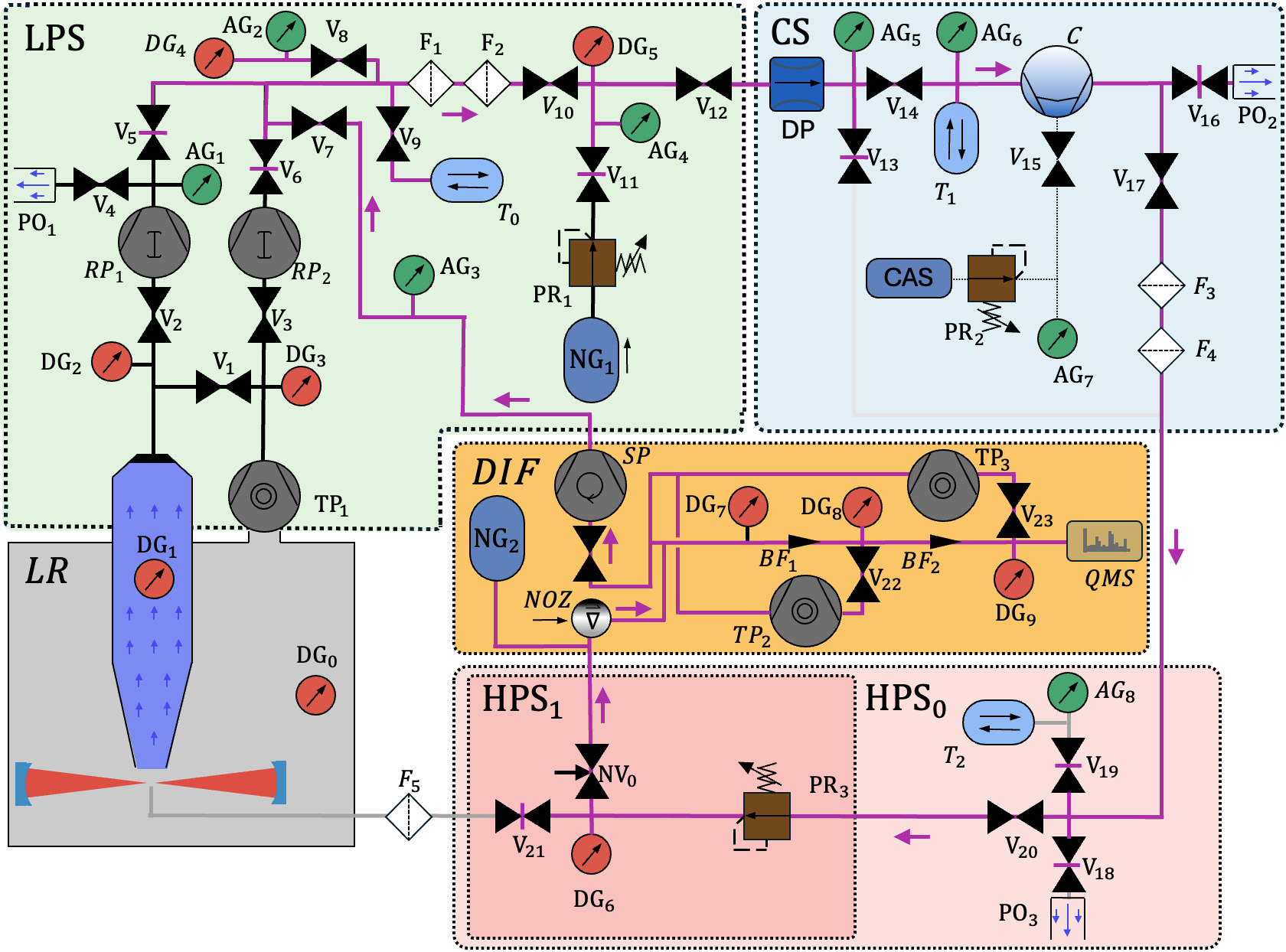}
    \caption{Experimental setup with all its modular components: Low pressure section (LPS), high pressure section (HPS), laser resonator (LR), compressor station (CS) and differential pumping (DIF) and the gas flow during the experiment (magenta lines). The DIF section is not strictly part of the system but used to test and quantify its operation. Valves marked with a magenta bar are closed. The nomenclature of the parts is shown in Table~\ref{tab:3.1.all_components}. The NG$_{x}$ (New Gas) ports are the locations were the system may be filled with new gas. NG$_{2}$ is mainly used for all measurements presented here as it allows for better monitoring of the pressures in the DIF section, as the DIF section is in another room from the rest of the system. NG$_{1}$ is used in normal operation.}
    \label{fig:Setup_gas_flow}
\end{figure*}

The assembled system in its current status is depicted in Fig.~\ref{fig:Setup_getrennt}. The layout is shown in Fig.~\ref{fig:Setup_gas_flow} and its components are listed in Tab.~\ref{tab:3.1.all_components} . The system is divided into four main sections, each highlighted in a different color: the low pressure system (LPS, green), scavenging the gas from the resonator and prefiltering it. The compressor station (CS, blue) raises the gas pressure to up to 60\,bar and does more filtering, and lastly the high pressure system (HPS, red), that monitors, controls and stabilizes the gas pressure. The laser resonator (LR) is shown in grey in the schematic overview and contains the gas catcher that constitutes the first part of the low pressure section.  The differential pumping stage (DIF, orange) is not part of the system itself, but it is essential as a test stand for the commissioning measurements reported in this work.

\begin{table}[h]
\centering
\caption{List of system components in the LPS, CS, HPS and DIF}
\small

\begin{tabular}{|l|>{\raggedright\arraybackslash}p{2.4cm}|>{\raggedright\arraybackslash}p{5cm}|}
\hline

\textbf{Abbr.} & \textbf{Name} & \textbf{Model} \\
\hline

\multicolumn{3}{|c|}{\textbf{Low Pressure System (LPS)}}\\
\hline

$AG_{2-4}$ & Analogue Gauge &
Manometers, 0 - 1000 mbar\\
\hline

$F_{1-2}$ & Filter &
Agilent Technologies - SCRINTRPNW25\\
& & Vacom - FT25KF-J\\
\hline

$T_0$ & Tank &
Festo - CRVZS-5\\
& & 5\,l Volume at 1 atm\\
\hline

$DG_{2,5}$ & Digital Gauge &
Balzers Compact Full Range Gauge\\
& & Telemecanique XMLPM05GD21 -1 - 5 bar\\
\hline

\multicolumn{3}{|c|}{\textbf{Compressor Station (CS)}}\\
\hline

$DP$ & Diaphragm Pump &
KNF-N630\\
\hline

$AG_{5-7}$ & Analogue Gauge &
Manometers, 0 - 1000 mbar \\
\hline

$C$ & Compressor &
Haskel - AG-15\\
\hline

$PR_2$ & Pressure Regulator &
FESTO - MS2-LR-QS6-D6-AR-MPA-B \\
\hline

$CAS$ & Compressed Air Supply & House supply - freeze dryed air\\
\hline

$T_1$ & Tank &
2\,L -- 16\,bar, RAL 5015\\
\hline

$F_{3-4}$ & Filter &
Balston - 95S6\\
& & Balston - E37/12\\
\hline

\multicolumn{3}{|c|}{\textbf{High Pressure System (HPS)}}\\
\hline

$PR_3$ & Pressure Regulator &
GCE Druva - CPLH0SJ \\ &&  Hastelloy membrane DruvaPUR 0 - 100 bar\\
\hline

$DG_6$ & Digital Gauge &
Telemecanique XMLG100D21\\
& & 0 to 100 bar\\
\hline

$NV_0$ & Needle Valve &
Swagelok SS-6BG-MM\\
\hline

\multicolumn{3}{|c|}{\textbf{Differential Pumping System (DIF)}}\\
\hline

$NOZ$ & Nozzle &
50\,$\mu$m diameter\\
& & Material: Borosilicate glass\\
& & Fraunhofer ILT\\
\hline

$SP$ & Scroll Pump &
Edwards XDS10\\
& & Dry scroll pump\\
\hline

$BF_{1,2}$ & Blank Flanges &
Custom-built\\
& & with 1\,mm hole diameter\\
\hline

$TP_{3,4}$ & Turbomolecular Pump &
Agilent TwisTorr 84 FS; Leybold TURBOVAC 50 (Art.-No. 85400)\\
\hline

$DG_{7-9}$ & Digital Pressure Gauge &
Balzers Compact Full Range Gauge; Balzers C.C. Gauge\\
\hline

$QMS$ & Mass Spectrometer &
Hiden HAL 201; RGA Controller; HTY18 Quadrupole\\
\hline

\end{tabular}
\label{tab:3.1.all_components}
\end{table}


Given that HHG is a non-linear process occurring in gas illuminated by high-intensity laser light, plasma unavoidably forms in the jet. Due to the high laser pulse repetition rate typically used in high-harmonic generation (for this setup, 40 MHz \cite{2_2_Review_50_years_Thirolf}), plasma generated by one pulse remains in the focal volume when the following pulses arrive. The accumulated ionized gas exhibits strong dispersion, which renders phase-matching of the harmonics impossible, leading to a significant loss in output power. Reduction of this effect is possible by ejecting the gas with a high pressure of 6-30 bar through a nozzle, creating a supersonic jet (here, a 50\,$\mathrm{\mu}$m diameter nozzle made out of borosilicate glass, the same type and size of nozzle that is used in the VUV-frequency comb\cite{Wissenberg2025}). 

The ejected gas is captured $\approx$ 0.8\,mm behind the laser beam by a conical "gas catcher" of 2\,mm opening diameter (illustrated by the blue volume in the grey section labelled "LR", in Fig.~\ref{fig:Setup_gas_flow}). A roots pump ($\mathrm{RP_{1}}$), placed further downstream in the low pressure section (highlighted in green in Fig.~\ref{fig:Setup_gas_flow}), creates sufficient gas flow so that no return flow occurs into the resonator. To maintain high vacuum conditions within the resonator chamber, a turbomolecular pump ($\mathrm{TP}$), backed by another roots pump ($\mathrm{RP_{2}}$), is employed.
To tackle possible mirror degradation, dominantly caused by hydrocarbons in the residual gas adhering to the mirror surface, multiple filters in the gases' path are utilized.
The first set of filters is located in the low pressure system, an adsorption trap and a HEPA (high efficiency particulate air) filter, (F$_1$ and F$_2$ in Fig.~\ref{fig:Setup_gas_flow}).
The zeolite adsorption filter (FT25KF-J, VACOM Vakuum Komponenten \& Messtechnik GmbH, Germany) removes hydrocarbons and  water vapor. An integrated heater regenerates the filter via baking out under vacuum (at  T $\approx$ 120\,$\degree$C). A subsequent HEPA filter (SCRINTRPNW25, Agilent Technologies Inc., USA) removes particles larger than 0.3~$\mathrm{\mu}$m with 99.97\,\% efficiency (possibly originating from the zeolite filter or abrasion from the scroll pump) to protect the diaphragm pump and compressor from mechanical abrasion \cite{AgilentDryScrollCatalog2023}.  

The compressor station (CS - blue color in the figure) raises the initial pressure from a few hundred millibar up to 30 bar (highlighted in blue in Fig.~\ref{fig:Setup_gas_flow}, constructed to our specifications by HD-tech Hochdrucktechnik GmbH \& Co. KG, Germany): Initially, a two-head sequential diaphragm pump reaches an intermediate pressures of 4 to 6 bar. All internal connections and components contacting the gas are lubrication free and stainless-steel. No contaminants can thus enter the gas stream. An intermediate tank with a volume of 2 L in the compression station stores gas at this pressure to dampen any pressure fluctuations caused by the next step.
A pneumatic compressor increases the pressure to the desired output value of 6 to 30 bar. Before entering the high-pressure system, the gas is filtered by F3 and F4 (95S6 and E37/12, Parker-Hannifin Corporation, USA) which remove hydrocarbons and particles down to 0.01\,$\mathrm{\mu}$m with up to $99.99999\,\%$ efficiency \cite{3_2_parker_balston_filters}. Any contamination entering the gas can therefore be attributed dominantly to  the tubing. These are all cleaned to the highest possible standards or in case of the tubing and pressure regulation purchased as ultra-high cleanliness components from the manufacturer. 

From the output of the compressor station, the high pressure gas can either be directed to the laser chamber through a 1/16" ultra-high purity swagelok line or stored in a tank T$_2$, enabling complete low-loss emptying of the system. If the gas is routed to the laser, a pressure regulator, PR$_3$, ensures a steady flow of gas with minimal pressure fluctuations at the nozzle. This pressure regulator (GCE DRUVA, Germany) sets the final nozzle pressure. Its effect will later be presented in Subsect.\ref{SubS:PressureStability}.
To ensure safe and stable operation of the gas system prior to installation on the laser resonator, a test stand was set up to replicate the nozzle setup and its surroundings under recirculating conditions.
The gas flow is highlighted in purple in Fig.~\ref{fig:Setup_gas_flow}; the test stand is coloured orange. 
Analysis of the residual gas components to check for hydrocarbons under recirculating operation is the central purpose of the test stand.
To detect these constituents, a residual-gas analyzer (RGA), i.e. a small quadrupole mass spectrometer (QMS) is used. 
As the QMS is only able to operate in pressures below $10^{-6}$~mbar total, differential pumping is employed. This allows continuous recirculating operation of the system with the nozzle directed at the QMS, but does not risk burning the filaments.
 A scroll pump reduces the pressure to $10^{-2}$~mbar in the main nozzle chamber. This is roughly the expected value in the resonator. To reduce the pressure further, but maintain directed gas flow to the QMS, two 1 mm apertures (BF1, BF2) and two turbomolecular pumps (TP2, TP3) form a two-stage differential pumping system, lowering the pressure to $10^{-6}$~mbar at the mass spectrometer and enabling continuous online measurements. The nozzle points at these openings.\\
With all pumps and the compressor running, this allows circulatory operation at variable pressures and simultaneous surveillance of all system parameters in real time.
Prior to first operation, the system was filled with helium and a leak tester with helium sniffer, which is calibrated against the background, was used to detect any leaks. After re-tightening all leaking connections, no measurable deviation from the background level was observed for the full setup. All chambers were tested using normal evacuated leak tester operation.

\section{System Operation} \label{Operation of the system}
Standardized procedures to operate the system in its different modes have been developed. The different modes, as well as startup and shutdown, are explained in this section.

\subsection{Starting the System}
Starting off, the system is assumed to be at ambient pressure and filled with air. 
First, V$_{10}$ is closed and a roughing pump is attached in place of T$_{0}$ on V$_{9}$ . Then the heating of the zeolite trap (Filter F$_{2}$) is initiated. This section of the system is pumped until the vacuum pressure has reached equilibrium, usually after roughly 12\,h. Then a small turbomolecular pump (Agilent TwisTorr 74FS) is attached at the same port and V$_{10}$ is opened again. The whole system is left with the circulation valves open and pumped for at least 48\,h, such that the 12\,m long pipes have sufficient time for outgassing. After a sufficiently low pressure (O(10$^{-6}$)\,mbar) is reached, V$_{9}$ is closed and the filling process can be initiated. 
Gas is filled into the system via inlet NG$_2$ or directly through the nozzle connection allowing for instant use of the laser. The roughing pumps in the low-pressure section are started first. When roughly 800\,mbar in front of the diaphragm pump (DP) are reached, this pump is also engaged. 6\,bar of pressure is built up in the intermediate reservoir T$_{1}$. Once this is reached, the compressor C may be engaged if higher pressures are necessary. The output valve V$_{17}$ is opened and the pressure stabilized. At this point the gas supply is shut off and the inlet NG$_2$ is closed. The system is now pressurized and ready for operation. Once the nozzle feed valve in the pressure regulator PR$_{3}$ is opened, gas will start circulating.

\subsection{Operation under Load}

When the system is to be operated under load, the only vitally important point is that the scroll and diaphragm pumps need to be activated before the nozzle feed valve is opened to avoid damage to the pumps or over-pressurization of the test/resonator volume.
Since VUV-laser stability strongly depends on a stable high-pressure gas supply at the nozzle, a pressure regulator, PR$_3$, was integrated into the system on the high pressure side in front of the nozzle, ensuring a constant, but manually variable, outlet pressure. Once circulation has started the nozzle pressure is set using this regulator. 
It is advisable that the backing pressure of the regulator be at least 5\,bar higher than the desired nozzle pressure. This ensures proper performance and long-term stability.
The goal is to achieve pressure fluctuations of less than 5\% of the set pressure at the pressure regulator PR$_3$. Fluctuations on a short time scale (less than 7 minutes) are targeted to remain below 1\%. In the following, unregulated gas flow refers to a setup configuration where the pressure regulator PR$_3$ is deactivated, allowing gas to flow unhindered. The achieved results will be discussed in Sect.~\ref{SubS:PressureStability}.
Once the pressure is set and has stabilized, the system is fully operational.

\subsection{System Shutdown and Gas Storage}
When the system is to be shut down, two procedures may be followed depending on the duration of the shutdown. 
For short shutdowns, when leaving over night, the best way is to close the nozzle feed valve V$_{21}$ and wait for the scroll pump to drop to chamber pressure. Once this is reached, its valve is closed. After the diaphragm pump can no longer reduce the pressure in the low pressure section, valve V$_{12}$ is closed. The gas is now sealed in the high pressure lines and intermediate reservoir.
Then the compressor, diaphragm pump and scroll pump are successively shut off. Leakage rates originating from this section are discussed in  Sect.\ref{SubSubS:Static}. Since they are sufficiently low, worries about gas loss are not warranted. However, over long term shutdown, admixture of gases from the low-pressure section that leak in  may become problematic, unless there is an intermediate outgassing of the zeolite trap and pumpdown undertaken.
For long term storage it is advisable to transfer the gas back into the high-pressure cylinder $T_2$. There, no low-pressure leakage into the system is relevant. This can also be used to temporarily store a gas species, without wasting it, while running the system with a different gas species.

\section{System Characterization and Monitoring}
To fully characterize the recycling system, it is essential to first identify the system’s requirements and then determine which physical quantities correspond to these requirements and how they can be experimentally measured.
This section sets out these requirements and how they are subsequently quantified and monitored in operation.

Given the high cost of xenon, the system must be as leak tight as possible. This not only accounts for expensive gas leaking out of the system, but also for external gases such as air leaking in. Any external gases not only make high-harmonic generation more inefficient due to unwanted plasma formation, but also introduce possible contaminants. Even a few monolayers of hydrocarbons can critically degrade the performance of an optical surface \cite{2_6_3_Hydrocarbons_VUV_Monolayer}, substantially reducing output power in the UV. 
As the nozzle pressure and thus gas jet speed are an important tuning parameter in HHG performance, pressure stability at the nozzle is of high importance to allow for stable operation.
To account for each requirement, different measuring instruments have been implemented in the system, which will be explained in the following, including related measurement procedures.

\subsubsection{Quadrupole Mass Spectrometer}
\label{MassSpectrometer}
The quadrupole mass spectrometer (QMS) is fed with a small portion of the direct jet flow in the test stand, allowing live monitoring of the gas composition. 
Previous tests showed that with the nozzle facing away from the spectrometer the results were altered slightly. With the jet facing the spectrometer, given the supersonic expansion into vacuum it is assumed that the differential pumping does not influence the relative jet composition due to the constituents' momentum. It just reduces the absolute pressure.
Two points of interest can be assessed with the QMS: hydrocarbon accumulation and leakage into the system contaminating the gas mixture.\\

Hydrocarbons cause mirror surface degradation, reducing the laser’s efficiency and power.
To detect hydrocarbons, partial pressures of gas species with a mass to charge ratio from 1 to 140 were recorded. This threshold was chosen as no gas species above a mass to charge ratio of 140 were detected within the mass spectrometer's sensitivity range. A pressure range from $1.33\cdot 10^{-9}$\,mbar to $1.33\cdot 10^{-8}$\,mbar was selected as this is most suitable for detecting the individual hydrocarbon contamination constituents listed in 12. Constituents above this pressure range were not observed. To generate the reference value by which to judge the overall gas quality, all components from this list were summed up, serving as a very sensitive upper boundary. 
Even though not a hydrocarbon itself, CO$_{2}$ is included in this accumulated contamination ledger as any carbon containing contamination was deemed as possibly negative for the mirrors at the expected laser intensities of the kW NIR and $\mathrm{\mu}$W VUV beams in the system, that are both in ranges where degradation is known to occur. 
Components with lower partial pressures than $10^{-9}$\,mbar were not detected. \\

The second point, leakage into the system, is investigated with a built-in table function, which identifies the seven most prominent residual gases and outputs their values iteratively over time. In the spectrometer, the main pumped gas has typical partial pressures of about $1\cdot10^{-6}$\,mbar. In the case of the testing measurements, the main gas was argon. Other dominant gases such as nitrogen, oxygen and water have partial pressures more than 26 times lower over the measurement duration. The measurements were taken multiple weeks after the last system bake-out and pump-down to create a worst-case operating condition and thus a hard upper boundary. Worst-case is used in the sense of negligent use without the maintenance necessary to suppress influence from leakage into the system, i.e. mainly water and hydrogen accumulating in the zeolite trap and the pipe walls, not system malfunction. 

The quadrupole mass spectrometer (QMS) is a refurbished HIDEN Analytical HAL 201. Its ion-source filaments were exchanged for fresh ones (new production, not new-old stock) and the QMS head was thoroughly cleaned and underwent heated vacuum drying. The back-end controller received a custom interfacing script allowing on-the-fly readout via a RS232 connection. The QMS heads were tested for radioactive contamination which was not present.
As very accurate knowledge about the present hydrocarbon contaminations was of paramount concern, the performance and lack of internal contamination was tested in an ultra-high vacuum environment. There proved to be no remaining contaminations to a $10^{-11}$\,mbar level. The hydrocarbon reference list utilized for this work can be found in Ref. \citenum{3_2_2_Hydrocarbon_fragments}.

\subsubsection{Pressure Gauges}
Pressure analysis provides information about leakage rate and gas flow stability of the recycling system.
\\ Two types of leakages can occur: Atmospheric air can flow into the system via the low-pressure system and valuable noble gas can leak out of the high-pressure section. As previously mentioned, analysis of the gases constituents with the mass spectrometer assesses whether air is leaking into the system. When the system is in circulating operation, it may still take some time to reach a steady state between the high-pressure and low-pressure system. The data of two pressure sensors is logged to take this into account:  DG$_{5}$ operating up to 5 bar for the low-pressure section and DG$_{6}$ for the high-pressure system with a measurement range of up to 100 bar (Pressure sensors XMLG100D21 and XMLPM05GD21F, Telemecanique, France). An ambient-pressure sensor is used as a reference. Furthermore, multiple analog pressure gauges (denoted in Fig.~\ref{fig:Setup_gas_flow} with AG) permit fast monitoring of the recycling system. This is especially useful in the startup and shutdown phases.
A steady flow of gas into the laser focus is necessary for stable HHG. Pressure oscillations, caused by the pumps in the system, may hinder this. Therefore, measures such as the buffer gas tank and the pressure regulator are employed and analyzed via these sensors. The observed oscillations occur on two distinct timescales, one on the scale of a few hours and one on the scale of a few minutes. Compared to that, pressure drops due to leakages take days. Thus they can be judged independently of one another.
\\Temperature fluctuations may also influence system pressure. However, measurements with a climate data logger (BL30, Trotec GmbH,  Germany) for more than 14 days showed that these influences are negligible. The laboratory temperature stability is sufficient with no more than $\pm 0.5\,\degree$C fluctuation over the whole measurement duration.

\section{Results}

Based on the parameters presented above, the final performance of the system is presented. 
During all measurements argon gas was used as the main jet gas. It is far less expensive and commercially more easily available than krypton or xenon, making it more suitable for loss assessment of the recycling system. Given xenon’s higher atomic mass and larger size compared to argon, the leakage rates, under final operation conditions with xenon, are expected to fall below the presented upper-bound estimate, as heavier and larger atoms exhibit reduced transmission through fissures in the system as well as lower cycle count through the system, as the nozzle consumption will also decrease. The improvement this brings may however be at least partially compensated by the reduced overall number of xenon atoms present in the system at the same pressure.

First, the pressure stability is analysed, and measures to optimize pressure stability are evaluated. This information is crucial for tracing possible reasons for laser power instability. Secondly, recovery efficiency of the gas is treated. Lastly, the results of long timescale mass spectrometry of the injected gas are presented.

\subsection{Pressure Stability}
\label{SubS:PressureStability}
Maximally consistent gas flow is required for stable high-harmonic generation and system pressure fluctuations are to be suppressed as much as possible. 
Testing revealed two types of oscillations, with periods of approximately 1 minute and 30 minutes, respectively. The oscillations can be attributed to the compressor C, which is powered by compressed air.
As the compressor increases the output pressure by a factor of 15 relative to the pressure of the incoming compressed air, small fluctuations in the supply lead to amplified changes in the output pressure and thus of the pumped volume. We attribute the 30 minute fluctuations to this effect, as this is roughly the frequency of the house compressed air system compressors turning on.
As the gradient of pumping speed vs. inlet pressure of the diaphragm pump feeding the compressor C is very steep, such effects are further amplified in the LPS. The 1 minute fluctuations are caused by the piston action of the compressor C.

Two measures aimed at minimizing the amplitude of these oscillations are separately discussed: the installation of a buffer gas tank T$_0$ (Fig.~\ref{fig:Fluctuations}) upstream of the diaphragm pump and the use of a pressure regulator PR$_3$ (Fig.~\ref{fig:Regulated}). The routing of the gas is marked in magenta lines in Fig.~\ref{fig:Setup_gas_flow}.

\subsubsection{Tank Connected vs. disconnected}
\begin{figure}[h]
    \centering
    \includegraphics[width=1.0\linewidth]{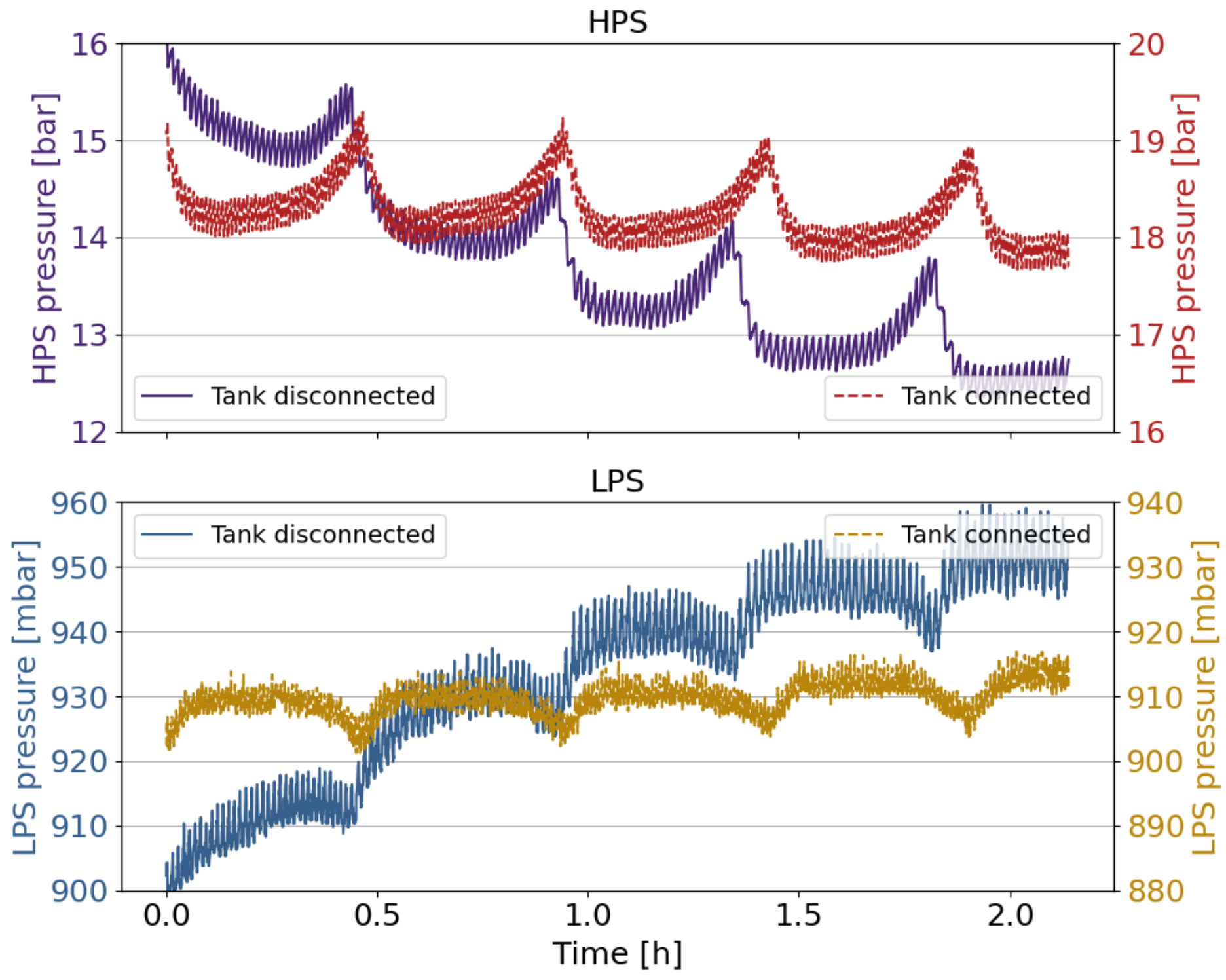}
    \caption{Comparison of pressure fluctuations and average stability with the buffer tank T$_0$ connected and disconnected. The scaling is consistent. Note that the traces with the buffer tank connected have both reduced fluctuations and a reduced gradient.}
    \label{fig:Fluctuations}
\end{figure}
Figure \ref{fig:Fluctuations} depicts the pressure oscillations with a connected and disconnected tank T$_0$. As the buffer gas tank is installed in the low-pressure section of the system and the gas flow to the high-pressure section is generated by the two-stage pumping system of the compressor station, no difference between a connected and disconnected tank scenario was found in the high-pressure section oscillation amplitude. However, in the LPS, the amplitude of the half-hour oscillation can be decreased by 60\% when a tank is connected. Likewise, the oscillations on a short timescale are reduced from 13-20 mbar to approximately 7 mbar. All in all, this corresponds to 1.8\% relative fluctuations (setpoint: 925 mbar) and 0.5\% for a connected buffer gas tank, which is within the chosen target margin of $\leq$\,1\%. 

The no-tank data in Fig. \ref{fig:Fluctuations} reveals an increase of the LPS pressure from 910 to 950 mbar during the two hours of measurement. As there is a concurrent decrease in the high-pressure section, this indicates that there is slower system equilibration without the tank.

\subsubsection{Regulated vs. Unregulated Gas Flow}

\begin{figure}[h]
    \centering
    \includegraphics[width=1\linewidth]{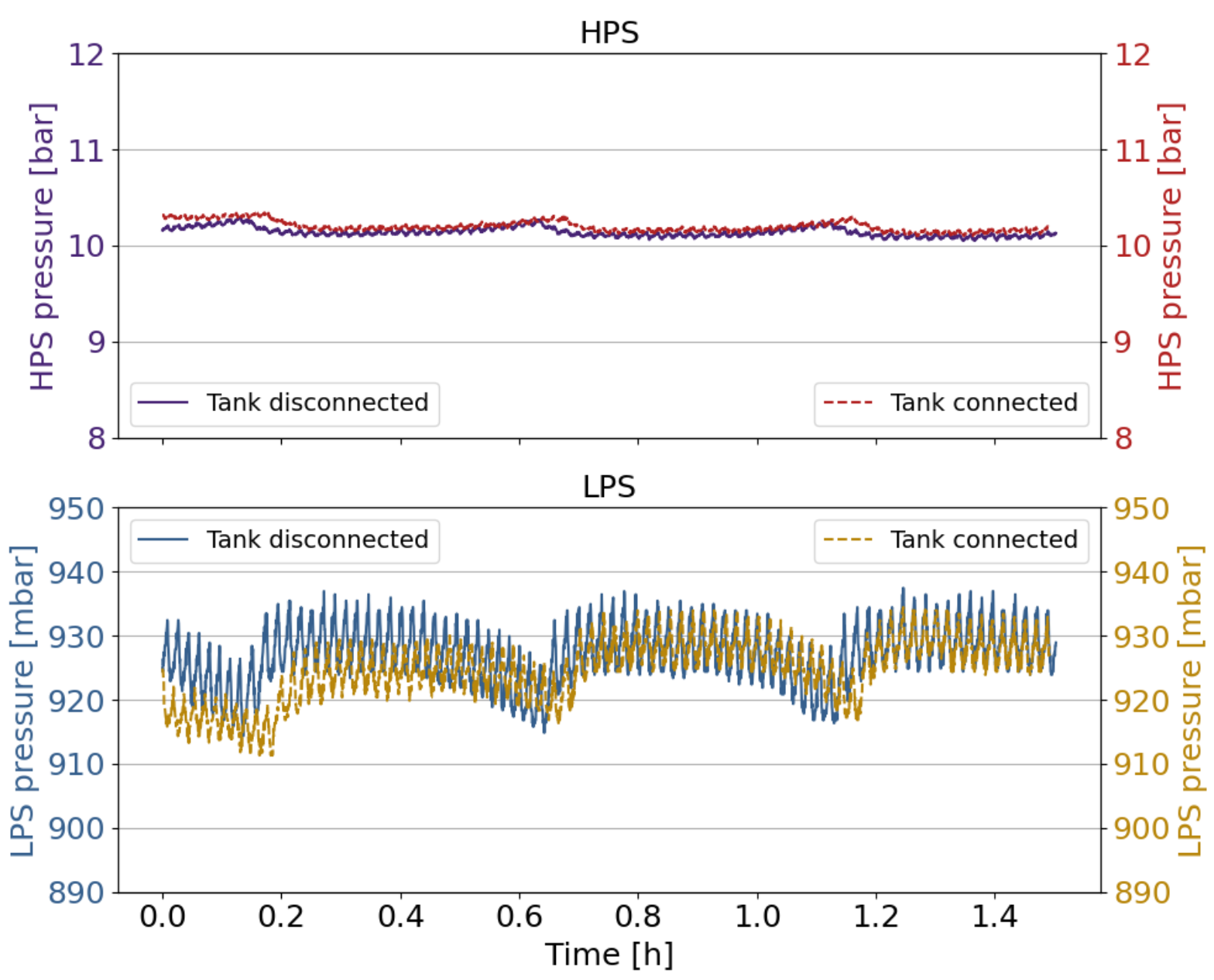}
    \caption{Pressure behaviour with activated gas pressure regulator. A clear difference in the high pressure
section is observed when compared to the unregulated gas flow in Fig.~\ref{fig:Fluctuations}. Usage of a pressure regulator
significantly reduces the fluctuations. Both Figure \ref{fig:Fluctuations} and \ref{fig:Regulated} use the same scale for comparison sake to clearly show the relative improvements.}
    \label{fig:Regulated}
\end{figure}
Comparing  Fig.~\ref{fig:Fluctuations} and Fig.~\ref{fig:Regulated} it is directly evident that with regulator PR$_3$ activated, fluctuations are reduced from about 4 bar to 0.25 bar - a 16-fold decrease in maximum fluctuation amplitude. On a short 7-minute scale, the pressure fluctuation amplitude feeding the nozzle is decreased from approximately 0.4 to below 0.06 bar. This corresponds to relative fluctuations of 0.5\% for short-time fluctuations and 2.5\% relative fluctuations (setpoint: 10.2 bar) for long-time fluctuations. 
As the pressure regulator is placed in the HPS in front of the gas nozzle, no significant pressure change is observed in the LPS when using it. 
This impact of regulating the gas flow is discussed when the tank is connected to the system. Measurements with a disconnected tank yield the same results for the amplitude of the pressure oscillations in the HPS.

\subsection{Recovery Efficiency}
Xenon will be kept in the system under two conditions: Either the laser is in operation and a continuous gas flow is maintained or the system is shut down and the valuable noble gas remains stored within the pipes. To replicate these scenarios, gas was routed on the same path as in the pressure fluctuation series, marked with magenta lines in Fig.~\ref{fig:Setup_gas_flow}. To assess the time until one standard liter of gas is lost, which corresponds to  N$_{st}=2.69\cdot 10^{22}$ particles at standard conditions (1 atm pressure~=~1013.25 mbar and 273.15 K), the ideal gas law can be rewritten: 
\begin{equation}
    N_{st}=\Delta N=\Delta p \frac{V_{system}}{k_B T} \Longrightarrow \Delta t=\left(\frac{\Delta p}{\Delta t}\right)^{-1}\cdot \frac{k_BTN_{st}}{V_{system}}\;.
    \label{eq:time_standard_liter}
\end{equation}
The laboratory temperature T is constant at 23$^{\circ}$C (=~296.15 K). The volumes V$_{system}$ are ($0.17\pm0.01$) liters in the HPS and ($9.32\pm0.05$) liters in the LPS, respectively. The change of pressure over time $\frac{\Delta p}{\Delta t}$ is logged by the pressure gauges.
The leakage rate I$_V$ (in mbar l/s) is a quantity typically used to describe how fast gas escapes from the system \cite{2_6_2_3_Flow_types_Leakage_Rates_Papa_Buch}:
\begin{equation}
    I_V=V\cdot \frac{\Delta p}{\Delta t}
    \label{eq:leakage_rates}
\end{equation}

\subsubsection{Static Measurements}
\label{SubSubS:Static}
\begin{figure}
    \centering
    \includegraphics[width=1\linewidth]{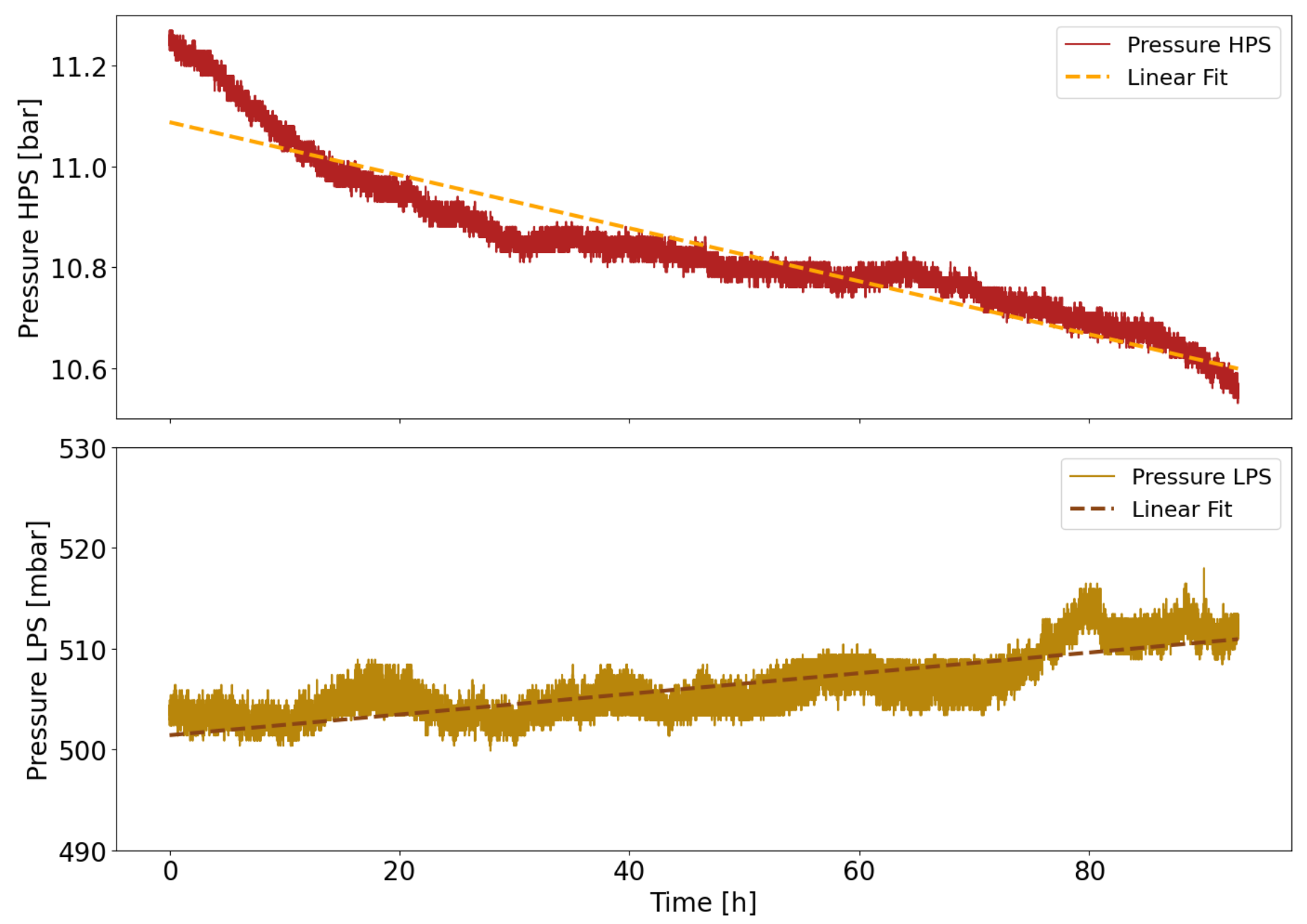}
    \caption{Pressure traces for the system pressurized but otherwise static, i.e. no flow. LPS and HPS are separated by closed valves. Thus, the HPS trace presents gas egress out of the system, the LPS trace ingress into the system.}
    \label{fig:Static}
\end{figure}

To determine how long xenon can be stored in the system without having to transfer it to a separate storage tank, static measurements were performed without gas flow. Because of the tiny nozzle
diameter (50 $\mathrm{\mu}$m), reaching a pressure equilibrium between all modular components takes multiple days. To allow for a more detailed investigation of individual leakage rates, the sections were therefore decoupled. All pumps were shut down and NV$_0$, V$_{12}$, V$_{17}$ were closed, isolating CS, LPS and HPS from each other for 92 hours. The drop in the high-pressure section, and corresponding small increase in the low-pressure section is clearly visible in Fig.~\ref{fig:Static}. This can be partially attributed to the compressor piston of C$_{0}$ slowly moving upwards and increasing the system volume as is also shown in Ref. \citenum{1_Griessbach_Paper}, where a similar compressor is utilised. This gives a lower bound for the leakage rate in the static case.
Fits yield the change of pressure over time, which is plugged into Eq.(\ref{eq:leakage_rates}) and Eq.(\ref{eq:time_standard_liter}) (listed in Tab.~\ref{tab:leakage_rates_static}).
\begin{table}[h!]
\centering
\caption{Leakage rates of the high-pressure section (HPS) and low-pressure section (LPS) are listed together with the time $\Delta t$ until one standard liter is lost in the high-pressure section and one liter of air sucked into the low-pressure section.}
\label{tab:leakage_rates_static}
\begin{tabular}{lcc}
\hline
Section & $\mathrm{I_V\ [mbar\; l/s]}$ & $\Delta t$ \\
\hline
LPS &  $(2.65 \pm 0.02)\cdot 10^{-4}$ & 50 days\\
HPS &  $(-2.41 \pm 0.01)\cdot 10^{-4}$ & 48 days\\
\hline
\end{tabular}
\end{table}

The calculated numbers in Tab.~\ref{tab:leakage_rates_static} show that approximately 50 days have to elapse before one standard liter of gas is lost from the HPS. On a similar scale, it takes roughly 48 days of one standard liter of air to be sucked into the LPS. For comparison: At typical operational settings, like in the here presented measurements, ($18.21\pm0.06$) standard liters of gas are enclosed in the entire system, which would be contaminated with about 0.1$\%$ of air
per day. Given xenon’s higher atomic mass and larger radius compared to the argon used in the measurements, the actual leakage rate is expected to fall below this upper-bound estimate as heavier and larger atoms exhibit reduced transmission through the micro leaks in the system.

\subsubsection{Continuous Gas Flow}

\begin{figure}[h]
    \centering
    \includegraphics[width=1\linewidth]{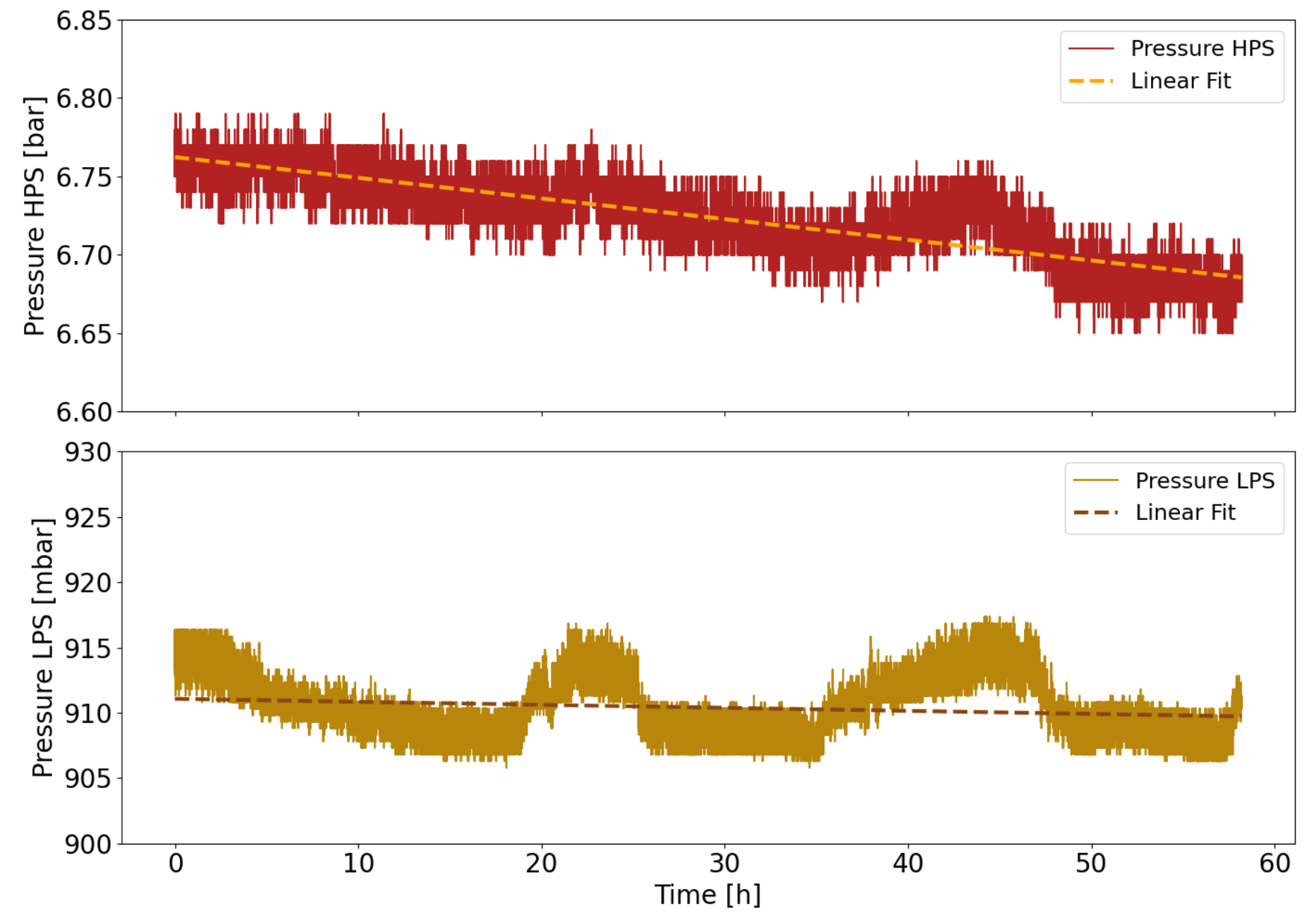}
    \caption{System pressure without pressure regulator. This is the only way in which a pressure drop in the HPS can be recorded, as with regulator applied, all changes would be compensated until the set pressure is reached. The compressor is off. The system pressure is generated only by the diaphragm pump.}
    \label{fig:6bar}
\end{figure}

\begin{figure}[h]
    \centering
    \includegraphics[width=1\linewidth]{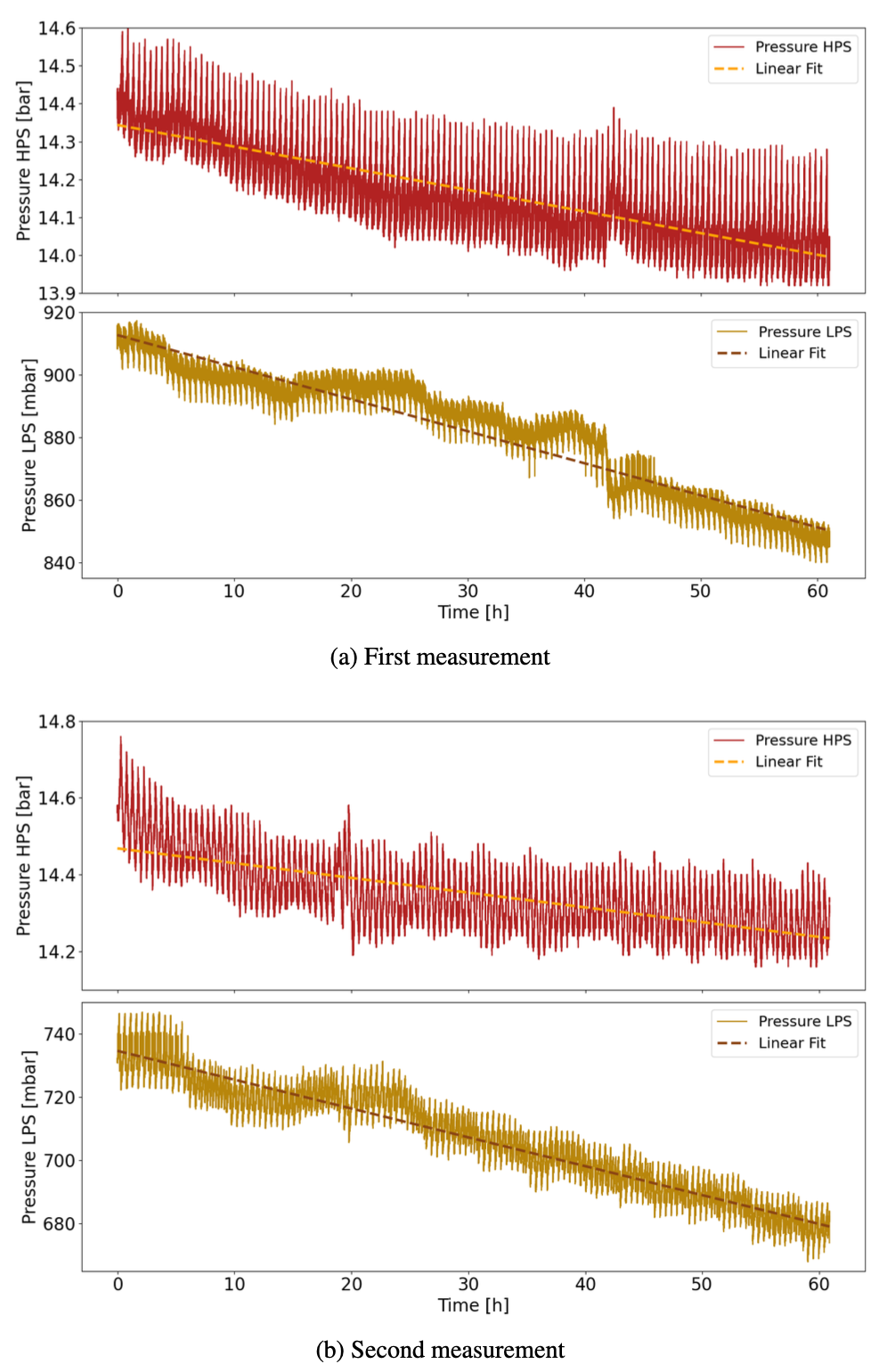}
    \caption{System pressure without pressure regulator, set to a nominal pressure of 14\,bar using the compressor. This is the only way in which a pressure drop in the HPS can be recorded, as with regulator applied, all changes would be compensated until the set pressure is reached. The compressor is off. }
    \label{fig:14bar}
\end{figure}

During operation, the gas jet parameters are predominantly determined by the pressure feeding the nozzle. As different operational schemes or gas mixtures require different pressures and flow rates, different cases are examined.
Leakage rates were measured at 6, 10 and 14 bar, which are likely operational pressures for the laser system to which this recycling setup will be connected. 
The pressure changes over time in the different sections were logged. The logs show that a pressure decrease in the HPS does not appear together with a corresponding increase in the LPS. This indicates that no transfer into different sections of the system is occurring, but that gas from the HPS is lost to the outside atmosphere. For the 6 bar measurement, the compressor C was switched off. As the compressor only uses blowback valves, gas flow in forward direction is possible even with the compressor not in operation.
As they are most representative of the different types of system behaviour, respectively, plots and discussion for the measurement with  6\,bar (Fig.\ref{fig:6bar}) and 14\,bar (Fig.\ref{fig:14bar}) are presented. The intermediate pressure of 10\,bar is only tabulated for completeness sake, as it is only different in values from the 14\,bar case where both the diaphragm pump and pneumatic compressor are running. The 6\,bar case is fundamentally different because the pneumatic compressor is not active.

\begin{table*}[]
\centering
\caption{Observed pressure change over time from the fit of different measurement series. From there, the leakage rate I$_V$, the time $\Delta t$ until one liter of gas is lost under continuous gas flow as well as the corresponding recycling efficiency $\eta$ of the system, are derived. It is important to note that this is the raw system efficiency and does not include losses possibly incurred by the gas catcher in the actual application supplying the VUV laser.}
\begin{tabular}{lccccc}
\hline
Measurement 
& $\Delta p / \Delta t_{\mathrm{HPS}}$ [mbar/s] & $I_V$ [mbar\,l/s] & $\Delta t$ [h] & $\eta$ \\ 
\hline
6 bar & $(0.37 \pm 0.01)\cdot 10^{-3}$ & $(0.98 \pm 0.02)\cdot 10^{-4}$ & 3122 
& 99.996\% \\

10 bar & $(1.72 \pm 0.01)\cdot 10^{-3}$ & $(15.90 \pm 0.16)\cdot 10^{-4}$ & 192 
& 99.936\% \\

14 bar -- 1st & $(1.58 \pm 0.01)\cdot 10^{-3}$ & $(42.9 \pm 0.4)\cdot 10^{-4}$ & 71 
& 99.829\% \\

14 bar -- 2nd & $(1.07 \pm 0.01)\cdot 10^{-3}$ & $(38.1 \pm 0.4)\cdot 10^{-4}$ & 80 
& 99.848\% \\
\hline
\end{tabular}
\label{tab:continious_gas_flow}
\end{table*}

The ultimate performance parameter, i.e. the system recycling efficiency $\eta$, is given by:
\begin{equation}
    \eta=1-\frac{I_V}{Q_{noz}+I_V} ~,
\end{equation}
where I$_V$ are the leakage rates and Q$_{noz}$ is the flow through the nozzle in mbar·l/s. It indicates the amount of gas which is returned to the system after it has left the nozzle. This is the case for what is called "full closed-loop operation", i.e. that all pumps connected to the volume into which the gas is ejected are feeding back into the recovery system. 
Since no analytical model exists to determine the nozzle consumption, a value from a similar nozzle with d = 40\,$\mathrm{\mu}$m was used. Tests determined a nozzle flow rate of Q$_{noz}$ = 2.50 mbar·l/s \cite{4_1_Nele_BA}. In the actual experiment, a nozzle with a diameter of 50 $\mathrm{\mu}$m was used,
which has 56\% larger cross-sectional area and thus a higher gas consumption. This would increase the efficiency even further, as the calculations with the 40\,$\mathrm{\mu}$m nozzle underestimate the total efficiency. The values for the different operational pressures are shown in Table \ref{tab:continious_gas_flow}.
Even with this conservative assumption and the system being tested with argon, which is more mobile compared to xenon, the recycling efficiency does not fall below 99.8\%. For comparison, a maximum efficiency of 98\% was achieved in the only known comparable setup \cite{1_Griessbach_Paper} operated in a full-recycling mode at MPIK Heidelberg. This indicates a significant reduction in leakage volume. It is important to note, however, that since the publication of Ref. \citenum{4_1_Nele_BA} the Heidelberg system has been upgraded with a similar KNF diaphragm pump.

\newpage

\subsection{Gas Contamination}

\begin{figure}[h]
    \centering
    \includegraphics[width=1\linewidth]{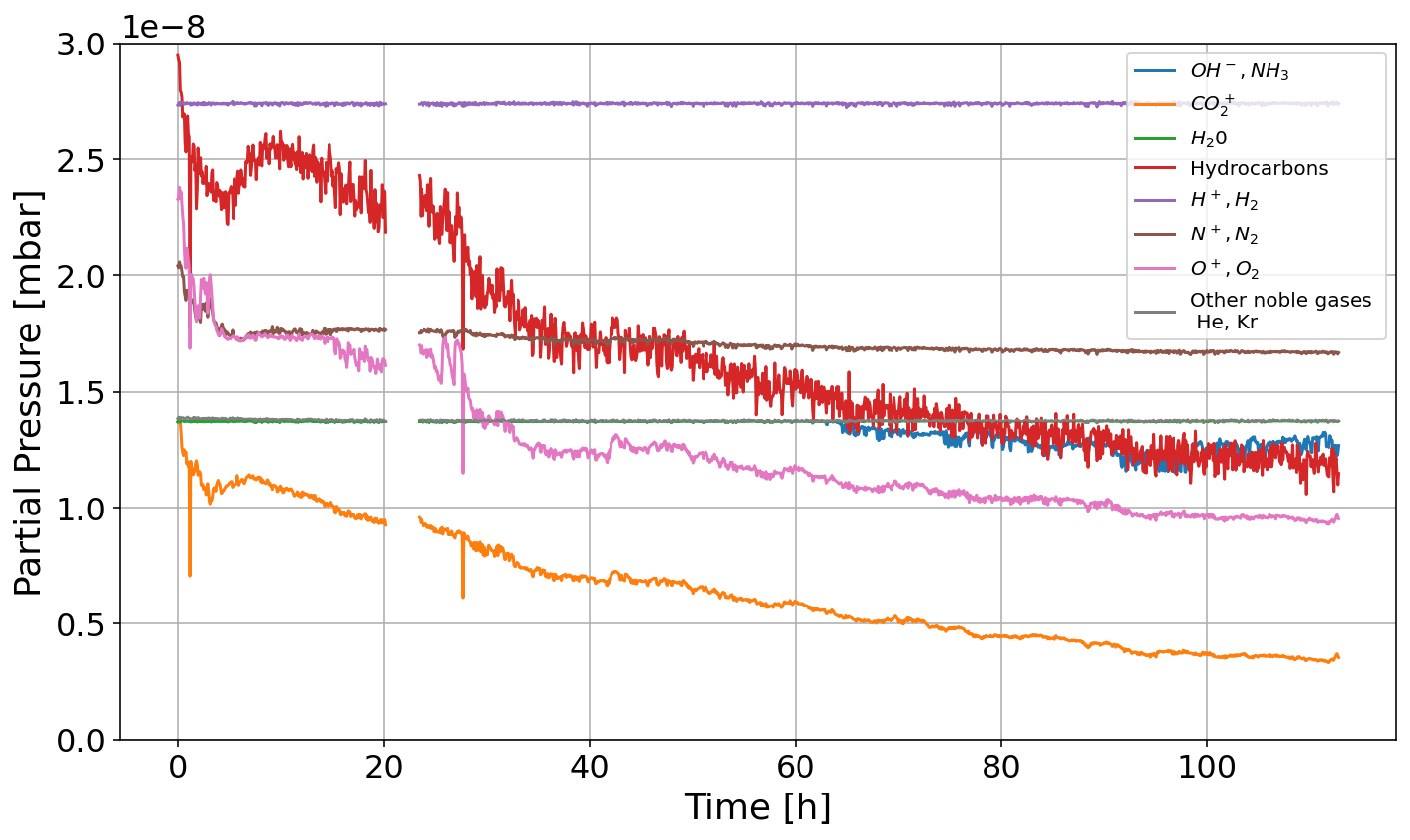}
    \caption{Temporal dependence of the partial pressures of different residual gas species with a fresh filling of argon circulating in the system at a nozzle pressure of 14\,bar: Once the partial pressures of the relevant constituents have come into the 10$^{-8}$\,mbar range where recording occurs, the contaminants partial pressures continue to fall. This is followed by a short period of partial pressure increase as the mechanical parts of the system and QMS warm up. Then, a long steady decline of hydrocarbon contaminations even after 110 hours of measurement occurs, indicating effective and continuous removal of these contaminants by the system's filters.
    The missing section after 20 hours is due to a short mode switch in the mass spectrometer. The setting changed automatically back to the initial selection about two hours later. The data can be considered continuous as the operation of the QMS itself did not change, just the readout of the signal in this timeframe.}
    \label{fig:hydrcoarbons}
\end{figure}

The system was pumped down and hydrocarbon levels were recorded at minimum pressure. A negligible total background in the 10$^{-9}$ mbar range was detected. Subsequently, a series of measurements was taken with continuous gas flow. A typical temporal development of the hydrocarbon levels is depicted in Fig.~\ref{fig:hydrcoarbons}. All measured data exhibits this progression over time: Remaining air is removed within the first hours of measurement, causing a rapid initial drop in all detected gas species. On longer time scales, the hydrocarbon level continues to decline as these species are trapped and removed, demonstrating the effectiveness of the adsorption traps and HEPA filters used. The decline of hydrocarbons settles at a  $\approx 1\times$ 10e$^{-8}$\,mbar level.
Other results such as Ref.\citenum{4_1_Nele_BA}, report hydrocarbon levels up to an order of magnitude lower. Their mass spectrometer, however, is installed in the large resonator vacuum chamber (approaching cubic metres of volume) connected to a multi-stage differential pumping system around the gas jet. This is a very high performance system and achieves very high throughput and efficient gas removal. Little direct gas flow is present near the mass spectrometer. In contrast, in the present setup with a small test chamber (V$\propto10^{-3}~m^3$) and fewer pumps, the total pressure at the mass spectrometer is higher under similar gas flow conditions. Consequently, the partial pressure of hydrocarbons and therefore the measured signal is also increased. Another difference may also be attributed to the method of data
evaluation: In Ref. \citenum{4_1_Nele_BA}, fewer hydrocarbon fragments were summed, as opposed to all stated in Ref. \citenum{3_2_2_Hydrocarbon_fragments}, however, the impact of this cannot be reliably quantified.
As stated above, it is also of note that our measurements were deliberately taken in what would be considered negligent operating conditions with weeks of system operation without the zeolite trap being heated and pumped. Therefore system performance in regards to gas contamination is considered to be sufficient to prevent damage to the laser resonator mirrors due to hydrocarbon contamination.
\\ In a further measurement series, the built-in table function of the QMS tracked the relative fraction of argon and air to the total gas mixture. No total decline over time was seen in a relevant order of magnitude.

\section{Conclusion}
The goal to build a compact and effective xenon recycling system, with a very high recovery efficiency of > 99,8\%, has been achieved. The desired smaller system size and complexity relative to the progenitor system at MPIK Heidelberg were accomplished and the design goals in terms of high efficiency, low contamination and stable gas quality were all reached.
Thus stable long-term operation of the VUV comb laser, crucial for applications, such as the realization of a nuclear clock, are possible.

\section*{Acknowledgemnts}

We are grateful to the staff of the Maier-Leibnitz-Laboratorium for giving us their HIDEN quadrupole mass spectrometer, as it is a vital system component. Furthermore, we thank our electronics engineer Dr. A. Zibell for helping service the mass spectrometer and constructing the pressure sensor readout electronics, as well as our local machine shop for all the needed fabrication work.

\section*{References}
\nocite{*}
\bibliography{Bib}

\newpage

\section*{Author Statements}
\subsection*{Conflicts of Interest}
The authors have no conflicts of interest to disclose.\\

\subsection*{Author Contributions}
G. Holthoff: Investigation, Conceptualisation, Methodology, Visualisation, Writing/Draft, Writing/Review \& Editing - lead,   Data Curation, Validation, Formal Analysis, Software - equal.\\
J. Rebhan: Software, Investigation - lead, Data Curation, Writing/Draft - equal.\\
T. Teuner: Investigation, Conceptualisation - equal, Data Curation - supporting.\\
L. Guth: Conceptualisation, Methodology - equal, Visualisation, Writing/Review \& Editing - supporting.\\
D. Moritz: Investigation - Supporting.\\
T. Teschler: Investigation - Supporting.\\
M. Wiesinger: Conceptualization, Formal Analysis, Investigation, Validation - equal, Writing/Review \& Editing - supporting. \\
J. R. Crespo López-Urrutia: Supervision, Conceptualisation,  Writing/Review \& Editing  - supporting.\\
P. G. Thirolf: Supervision \& Funding - lead, Validation, Writing/Review \& Editing - supporting.\\

\subsection*{Funding}
This work was supported by DFG (Th956/3-2) as well as by the European Unions Horizon 2020 research and innovation program under grant agreement 6674732 „nuClock“ and is currently supported by the ERC Synergy Grant „ThoriumNuclearClock“.

\subsection*{Data Availability Statement}

The data that support the findings of this study, as well as the corresponding code for its visualization/analysis, are available from the corresponding author upon reasonable request.

\end{document}